\documentclass[10pt,letterpaper]{article}
\usepackage[top=0.85in,left=1in,right=0.8in,footskip=0.75in]{geometry}

\usepackage{amsmath,amssymb}

\usepackage{changepage}

\usepackage{textcomp,marvosym}

\usepackage{cite}

\usepackage{booktabs}

\usepackage{nameref,hyperref}

\usepackage{fancyvrb}

\usepackage{longtable}

\usepackage[nopatch=eqnum]{microtype}
\DisableLigatures[f]{encoding = *, family = * }

\usepackage[table]{xcolor}

\usepackage{array}

\hypersetup{colorlinks=true, citecolor=blue, linkcolor=blue, urlcolor=blue}

\newcolumntype{+}{!{\vrule width 2pt}}

\newlength\savedwidth

\raggedright
\usepackage[aboveskip=8pt,labelfont=bf,labelsep=period,justification=raggedright,singlelinecheck=off]{caption}

\makeatletter
\renewcommand{\@biblabel}[1]{\quad#1.}
\makeatother

\usepackage{lastpage,fancyhdr,graphicx}
\usepackage{epstopdf}
\fancyheadoffset[L]{2.25in}
\fancyfootoffset[L]{2.25in}
\begin{document}
\vspace*{0.2in}

\begin{flushleft}
{\Large
\textbf\newline{Iridium: A Modern Graphical User Interface for Simulating SBML Models} % Please use "sentence case" for title and headings (capitalize only the first word in a title (or heading), the first word in a subtitle (or subheading), and any proper nouns).
}
\newline
\\
Herbert M. Sauro\textsuperscript{1,3},
Dan Nguyen    \textsuperscript{5},
Long Nguyenle  \textsuperscript{5},
Steven S. Andrews\textsuperscript{1},
Frank T. Bergmann\textsuperscript{4},
Joseph L. Hellerstein\textsuperscript{1,3,5},
Lucian P. Smith\textsuperscript{3},
H.\ Steven Wiley\textsuperscript{2}

\bigskip
\textbf{1} Department of Bioengineering, University of Washington, Seattle, 98195-5061, WA, USA

\textbf{2} Biological Sciences Division, Pacific Northwest National Laboratory, 902 Battelle Blvd, Richland, 99354, WA, USA

\textbf{3} eScience Institute, University of Washington, Seattle, 98195-5061, WA, USA

\textbf{4} BioQUANT/COS, Heidelberg University, INF 267, 69120 Heidelberg, Germany

\textbf{5} Allen School of Computer Science, University of Washington, Seattle, 98195-5061, WA, USA

\bigskip

% Use the asterisk to denote corresponding authorship and provide email address in note below.
Corresponding author: hsauro@uw.edu

\end{flushleft}
%\comment{jlh}{
%Joe's Overall Comments
%\begin{enumerate}
%    \item The paper has all of the right information. My focus was on making the motivation and impact more clear, mostly in the abstract and the conclusions, although I added some language elsewhere as well.
%    \item Should we reference Aldrich's JOSS paper for the rate checking feature?

 %   \item HMS: I don't use it because it wasn't comprehensive enough; we could cite it as an earlier attempt. 
%\end{enumerate}
%}

\section*{Abstract}

Biological models based on mathematical simulations (e.g., ordinary differential equations) are hugely beneficial for understanding biological processes and predicting the outcomes of experiments. Such models are built using a variety of sophisticated modeling tools. Unfortunately, the complexity of these tools is often a barrier to use by experimental biologists. To improve the usability and impact of biological models, we have created a desktop platform (DeskIridium) and a web-based simulator (WebIridium, hosted as a GitHub page) that, in combination, provide features that promote ease of use by experimental biologists, including: human-readable editing of models via the Antimony syntax; sliders for adjusting parameter values while simultaneously displaying simulation results in real time; zero-install distribution; AI chat integration; and generation of standalone Python implementations of SBML models to support reproducibility. The desktop version uses the well-established libroadrunner simulation package; the web-based application uses an Emscripten translated version of COPASI. Both simulators are SBML-compatible, allowing users to easily import previously built models or edit existing models using the easy-to-use Antimony language.
%We focus on usability and productivity, especially for novice users. Both platforms support interactive simulation and time-course simulations, steady-state analysis, parameter scanning, and sensitivity analysis, such as metabolic control analysis and the Jacobian.Multiple data files can be loaded into the simulator to compare directly with the simulation results and guide parameter estimation using interactive parameter controls. The simulators can export Python descriptions of simulation experiments to support reproducibility and portability. The web-based implementation has zero installation requirements, thus providing universal access to SBML-compliant models through a web browser. The Web version is hosted as a static GitHub page and can run on any modern browser, including on a tablet. These tools supplement Python-based scripting tools, such as Tellurium, by providing an easy modeling route for students and researchers who may not be familiar with Python.
We used modern AI methods for software development and share the lessons we learned. Binaries, source code, and the web interface are available at:

\noindent
\url{https://github.com/sys-bio/IridiumSimulator} and \url{https://github.com/sys-bio/WebIridium}.

%\linenumbers

% Use "Eq" instead of "Equation" for equation citations.
\section*{Introduction}

Biochemical systems are typically too complicated to understand without the aid of simulation tools. As a result, these tools have become essential to modern biological research.
%from simple reasoning, making dedicated simulation software an essential tool for modern research.
%Modeling biochemical systems is time-consuming and knowledge-intensive, requiring deep knowledge of biochemical pathways and the kinetic properties of individual steps. This has encouraged the development of software tools to provide assistance and as a result there is a long history of computer simulation tools of biochemical systems to simplify modeling~\cite{chance1962analogue}.
%Today's simulators build on a long history of prior work~\cite{chance1962analogue}.
The earliest is from 1943 when Britton Chance used the mechanical differential analyzer to model the action of the enzyme peroxidase~\cite{chance1943kinetics}. In 1952 Chance and coworkers used a custom-built electronic analog computer to model the action of catalase~\cite{chance1952mechanism}. From the late 1950s onward, however, the advent of the digital computer with its high speed and capacity coupled to potentially much greater numerical precision soon displaced the analog computer as the preferred medium, although analog computers were still used in the 1960s~\cite{chance1962analogue, higgins1964chemical}. By the 1980s the personal computer became a viable platform for running simulations~\cite{sauro1991scamp} and today, we have access to unprecedented computer power as well as rich user interface and graphical capabilities. In 2003, the first exchange format, SBML~\cite{hucka2003systems}, for biochemical models was released. This resulted in the emergence of large repositories of models such as BioModels~\cite{malik2020biomodels} where models are easily exchangeable and reusable.  Biochemical simulation applications are now commonplace and are available both as graphical user interfaces~\cite{hoops2006copasi,sauro2003next} and as packages for scripting languages such as Python~\cite{olivier2005modelling,somogyi2015libroadrunner,choi2018tellurium}. 

This paper introduces the Iridium simulation platform, a desktop simulator (DeskIridium) and a web-based simulator (WebIridium). This platform is specifically geared toward ease of use and rapid interactive operation. The objective in designing Iridium was to make computer simulation painless and productive. Many of the software tools available today require, in some cases, considerable setup and training. In designing Iridium, we partnered with seasoned modelers who require ease of use and high productivity. This is especially true when models become large. 

This article focuses primarily on the novelty of the human interface and the role modern AI can play in software development. For the simulation engines, we use existing tooling. On the desktop, Iridium uses the libroadrunner simulator library~\cite{somogyi2015libroadrunner,welsh2023libroadrunner}. On the Web, we use a JavaScript/WASM version of COPASI~\cite{hoops2006copasi}. The main novel contributions of the Iridium project are:
\begin{enumerate}
    \item Exploration of parameter values using sliders that provides rapid visualization of simulation results;
    \item Exporting models as Python implementations to promote reuse and reproducibility as well as SED-ML and COMBINE archives;
    \item A zero-install, universally accessible, and low-maintenance web-based implementation hosted on a static GitHub Page;
    \item AI chat that automatically includes model context to provide high-quality AI assistance for modeling, as well as built-in model checking in the desktop version;
    \item Rapid search and loading of the BioModels repository for models. 
\end{enumerate}

\section{Design Decisions}

As with all modern simulation platforms, the Iridium platform supports importing and exporting models using SBML. In addition, for reproducibility of simulations and analyses outside the tools, the desktop edition allows users to translate their workflow to an equivalent Python script, that can be run using the Tellurium Python package~\cite{medley2018tellurium,choi2018tellurium}. The desktop app can also export SED-ML and COMBINE archives.

\subsection{Guiding Principles}

Our guiding philosophy originates from a quote attributed to the computer scientist Alan Kay: `Simple things should be simple, complex things should be possible'. He coined this expression during his time at Xerox PARC  to emphasize that user interfaces and programming languages should be intuitive for beginners yet powerful enough for advanced tasks~\cite{XeroxParc}. Similarly, Larry Wall, the creator of the Perl programming language, expressed a similar sentiment: `Easy things should be easy, and hard things should be possible.' This philosophy guided Perl's design, aiming to make simple tasks straightforward while ensuring that more complex operations remained achievable~\cite{Perl}. The design of the Iridium platform was driven by these principles. 

\subsection{Panel Structure}

The application is a single window (or page on the Web version) that is divided into three panels. From the left, the panels include:

\begin{enumerate}
\item Control Panel
\item Input Panel
\item Output Panel
\end{enumerate}

\begin{figure}[htb]
\centering
\includegraphics[scale=0.165]{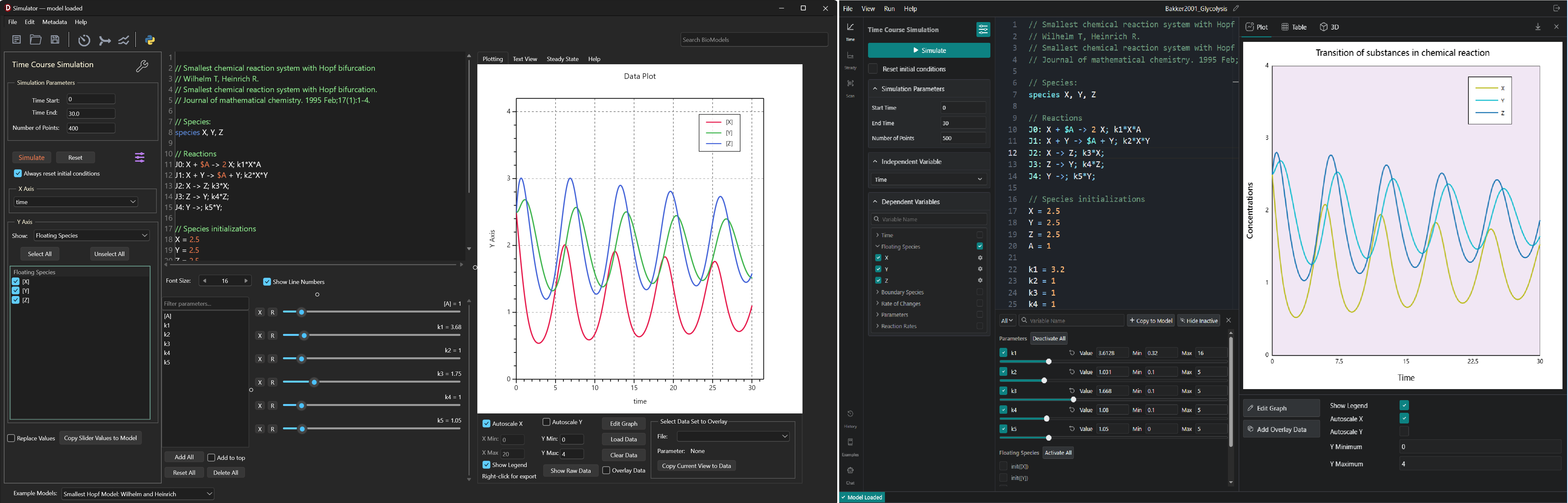}
\caption{Screenshots of the main window of the desktop version (left) and the Web version (right) of Iridium. To the left is the control panel, the center includes the models and other input features such as the sliders, and on the right is the output panel, which shows a 2D plot of a simulation.}
\label{fig1}
\end{figure}

The user can select different tools by switching the control panel via the toolbar. The current range of tools includes the following:
\begin{enumerate}
\item Time Course Simulation  % No stochastic?
\item Steady State
\item Parameter Scan
\end{enumerate}

\noindent
{\bf Time Course Panel:} The time course panel can be used to run deterministic simulations of the current model in the input panel. The duration of the simulation and the number of points to generate can be specified, as well as what simulation outputs should be captured. The panel allows a user to specify an independent variable and one or more dependent variables. By default, repeated calls to simulate will reset the initial conditions back to their original values. Alternatively, the panel can be configured so that a new simulation continues from where the last one finished. 

All input control panels have a slider option, which, when selected, creates a new slider panel in the central input panel (Figure~\ref{fig1}). This allows users to easily and rapidly interact with the model. Users can easily create or delete sliders and set their ranges. Users can copy slider positions back to the model if a particular behavior has been identified and needs to be captured in the Antimony script. 

\noindent
{\bf Steady State Panel:} The steady-state panel is used to compute the steady state of the current model. It will also compute the Jacobian and sensitivity coefficients defined in metabolic control analysis~\cite{fell1992metabolic,kacser1995control}, both scaled and unscaled sensitivities. The desktop version also offers a 3D plot of the sensitivities.

Results from the steady state analysis can be copied to the clipboard or saved as a CSV file. As with the time course panel, the steady state panel has a slider option. However, unlike the time course sliders, this does not allow the user to compute the steady state as the slider moves; instead, it computes the steady state only when the user releases the slider. This is because computing the steady state in real time cannot be guaranteed due to the nature of the algorithms employed. Instead, a separate tool has been developed to study how the steady state changes as a function of parameters, as well as to provide stability information. This is based on the Julia BifurcationKit package~\cite{BifurcationKitJulia,BifurcationGUIApp}.

\begin{figure}[htb]
\centering
\includegraphics[scale=0.3]{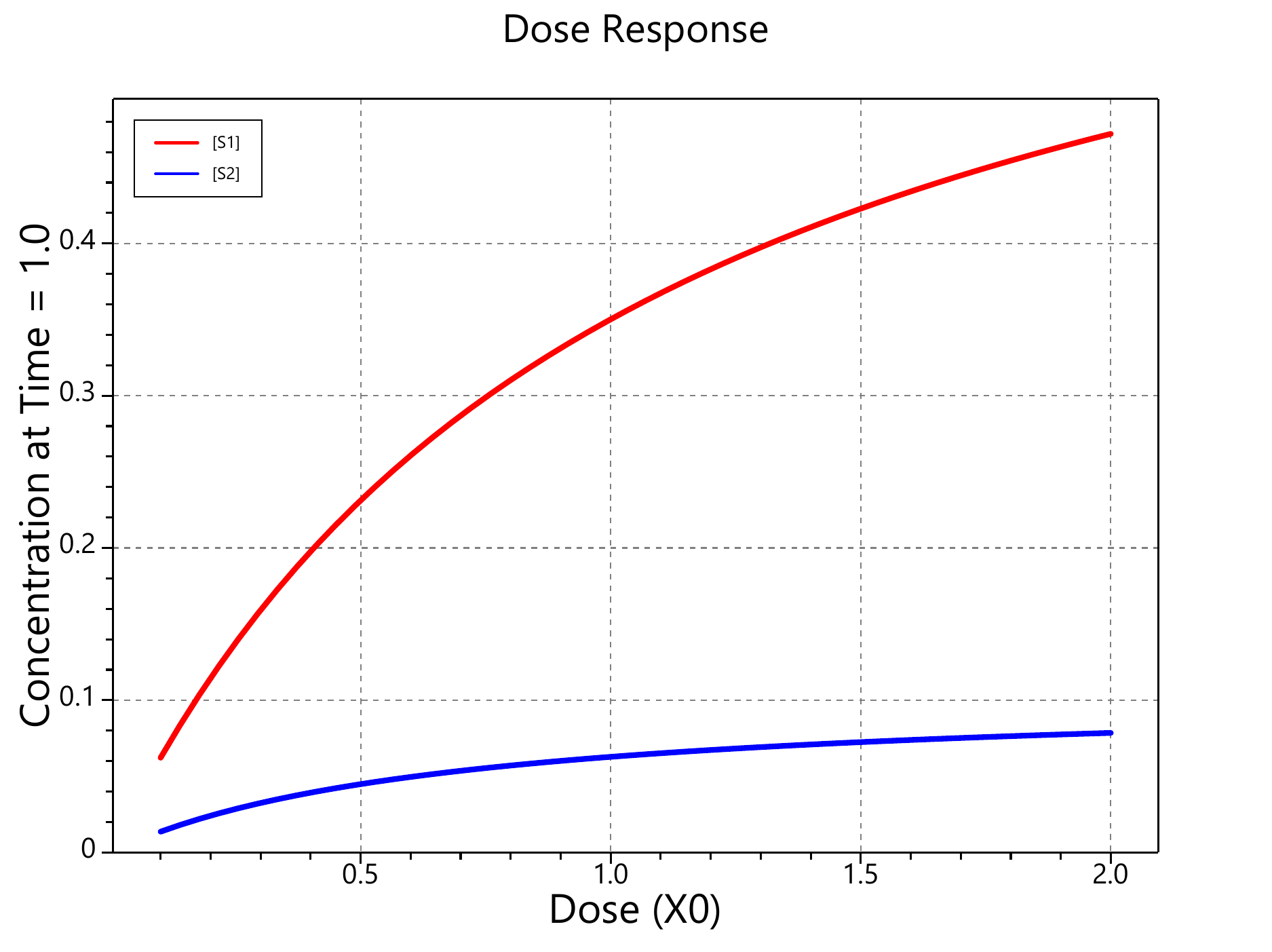}
\caption{Parameter scan used to trace the effect of dosing the input species, {\tt X0} and observing the levels of S1 and S2 at time equal to 1.0. The model comprised three steps {\tt X0 -> S1 -> S2 ->}, where {\tt X0} was a boundary species, with each step governed by a reversible Michaelis-Menten rate law.}
\label{fig:DoseScan}
\end{figure}

\begin{figure}[htb]
\centering
\includegraphics[scale=0.35]{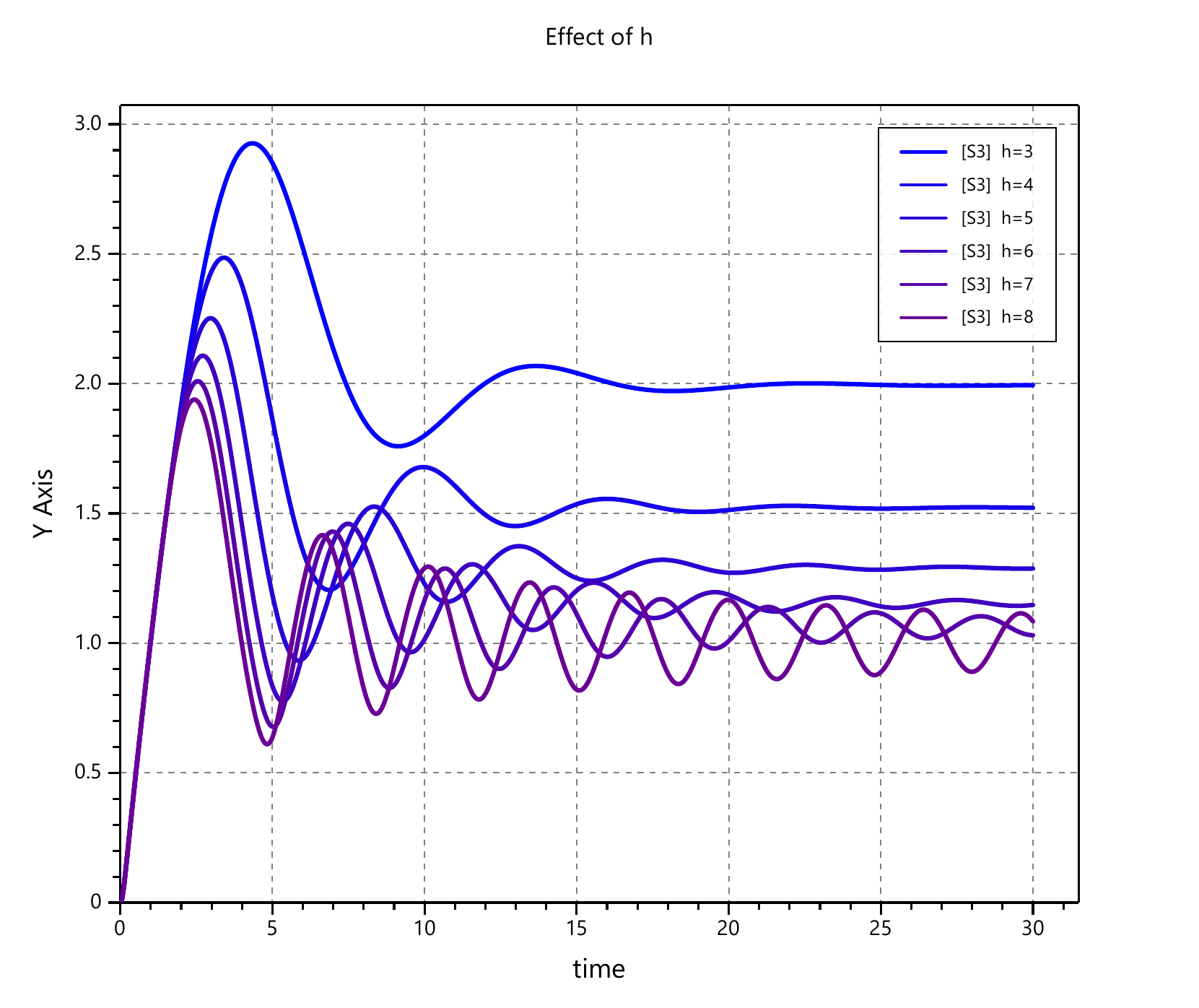}
\caption{Parameter scan for an oscillating system. The scan changes the model's Hill coefficient from 3 to 8 across six values. The scan illustrates the onset of oscillations as the feedback strength increases. The exported Python script is shown for illustration purposes in the Appendix.}
\label{fig:ParameterScan}
\end{figure}

\noindent
{\bf Parameter Scan Panel:} The parameter scan panel generates data of time courses or specific points in time as a function of changes in a parameter value (Figure~\ref{fig:ParameterScan}). Different model outputs can be selected, and one parameter can be selected for a given scan. Scans can be linear, log, or defined by manually assigned values in a list. Scans can be used to track the time course, endpoint, peak value, or time to peak as a function of a given parameter. Scans can also be used to examine the dose response by plotting the response at a given time as a function of dose (Figure~\ref{fig:DoseScan}).   

Sliders can also be used with a parameter scan. The only restriction is that the parameter being scanned doesn't have a slider; only non-scanned parameters have sliders. 

%\begin{figure}[htb]
%\centering
%\includegraphics[scale=0.5]{Iridium2.png}
%\caption{A close-up of the time course simulation control panel. Simulation %times, and output selections can be made with this panel to set the x and y %axes. A variety of outputs can be selected, including fluxes, rates of change, %elasticities, eigenvalues, etc. The panel also includes a slider button to %bring up the interactive modeling panel.}
%\label{fig:TimeCoursePanel}
%\end{figure}

%\begin{figure}[htb]
%\centering
%\includegraphics[scale=0.25]{Iridium3.png}
%\caption {Screenshot showing the steady-state control panel with the output window on the right showing the flux control coefficients.}
%\label{fig3}
%\end{figure}

{\bf Input Panel:} The central input panel has two halves. The upper half is an editor where a user can specify a model using the Antimony syntax. The editor provides line numbers for error reporting and syntax highlighting.

The lower panel is reserved for other input modalities. In the current version, the only other input mode is the slider panel.

\noindent
{\bf Output Panel:} Time course simulation results can be output in two forms: 2D plots or CSV tables. Steady state results are output either as tables or 3D bar plots. The 2D plotting control has a dedicated configuration panel that allows users to change colors and other stylistic aspects. All plots can be saved in the form of PDF files, suitable for publication. Bitmap PNG format can also be exported. Alternatively, a user can save the raw data as CSV and use their own plotting software. 

\begin{figure}[htb]
\centering
\includegraphics[scale=0.3]{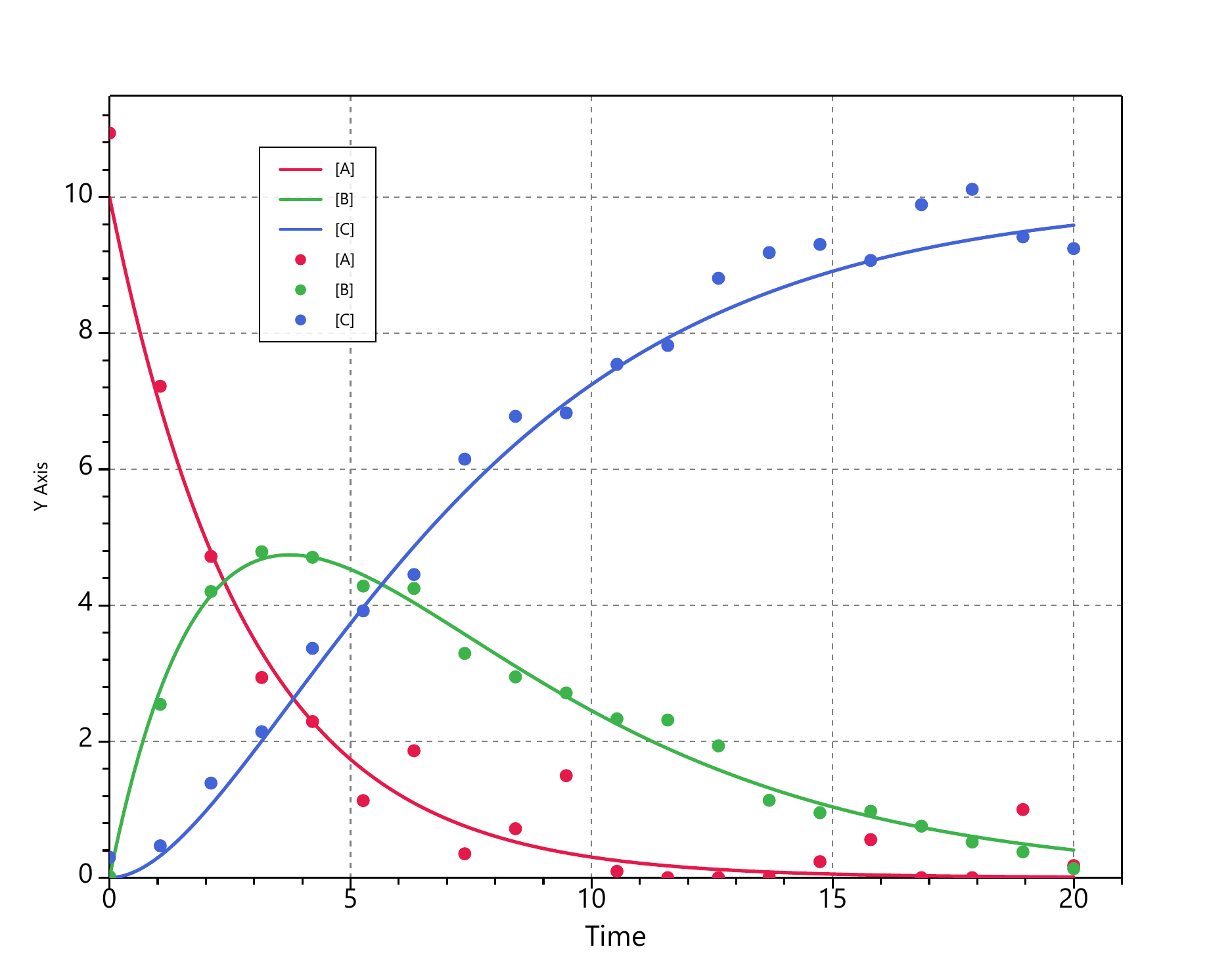}
\caption{Simulation showing data overlaid onto the results of the simulation.}
\label{fig:Dataoverlay}
\end{figure}

Data from an external source, in the form of CSV files, can be overlaid on a simulation study. This allows users to compare a time course simulation with experimental data. Any number of files can be loaded and overlaid or shown individually (Figure~\ref{fig:Dataoverlay}). 

\noindent
{\bf Model Checking:} The desktop version has a built-in model checking feature that examines models for errors or warnings such as suspicious rate laws or missing initializations of simulation variables. In WebIridium, the built-in AI assistant provides similar capabilities. The checker will also do dynamic tests and can check and identify, for example, divide-by-zero errors. Model checking is based on a series of eighteen built-in rate-law templates; these can be exported as JSON and modified, or new rate laws can be added by creating new JSON files.  A model-checking report is provided as a Markdown document in the reports tab of the output panel. This report, together with the Antimony model, can be provided to an AI to fix the identified issues in the model. The appendix contains an example of a report from the checker which has identified a suspicious rate law {\tt k1*A/Km} and a divide by zero error, since {\tt Km = 0}.

\subsection{Simulation Engines}

\subsubsection*{Desktop Version}

The underlying simulation engine for the desktop version is libroadrunner~\cite{somogyi2015libroadrunner}. This is a high-performance SBML-compliant simulator library with an extensive C and C++ API. The engine was designed to be hostable by other applications, and a variety of tools use it for this purpose~\cite{wikipediaLibroadrunnerWikipedia}. libroadrunner also has an extensive Python interface that allows users to script simulation studies within Python. Iridium uses the C API~\cite{CAPI} to gain access to the libroadrunner's functionality.

\subsubsection*{Web-based Version}

The web-based version uses COPASI \cite{hoops2006copasi} as its backend simulator, bringing a robust and SBML-compliant engine directly into the browser. While libroadrunner is the simulation engine used in the desktop edition, its just‑in‑time compilation approach is designed for native environments and is not currently compatible with browser execution. To enable fast, installation‑free simulations on the web, COPASI was compiled to WebAssembly (Wasm) \cite{w3cwasm} using Emscripten~\cite{Zakai2011Emscripten}, and a lightweight JavaScript interface provides convenient access to its functionality.

The remainder of the Web interface was developed using React and can therefore be run from a static web server such as GitHub. This makes deployment much easier and doesn't require an active server to operate. Hosting the web version is simply a matter of moving the JavaScript and Wasm files to a new host. 

\section{Modeling Standards}

The Iridium platform uses a variety of community standards. This makes it possible to exchange models and modeling experiments with other software platforms. We define the standards we use in Table~\ref{tbl:standards}.

\begin{table}
\centering
\begin{tabular}{lp{9cm}} \toprule
Standard & Purpose \\ \midrule
SBML~\cite{hucka2003systems} & Systems Biology Markup Language: This is a commonly used standard for exchanging systems biology models between simulation tools or for storing models in repositories such as BioModels. \\
SED-ML~\cite{waltemath2011reproducible} & Simulation Experiment Description Markup Language: This is a proposed standard for describing simulation experiments.  \\
COMBINE Archive~\cite{bergmann2014combine} & This is a zip file with a manifest that can be used to store the various components of a modeling project. For example, it can store the SBML model as well as the SED-ML file.  \\ \bottomrule
\end{tabular}
\caption{Modeling Standards used by the Iridium Platform}
\label{tbl:standards}
\end{table}

Both the desktop and web platforms support the import and export of SBML. The desktop version also supports the export of SED-ML and COMBINE archives. While SBML is used to describe the model, SED-ML is used to specify the simulation experiments done by the user.  This helps to ensure that a given simulation experiment is reproducible. Several tools support SED-ML, such as COPASI, Tellurium, and VCell, as well as non-SBML tools such as OpenCOR~\cite{garny2015opencor}.

To implement SED-ML support, we developed a simple experiment description specification format that can be embedded in a comment section. This structure can be converted to SED-ML. A more detailed description of this format is given in the appendix together with a sample script. Full documentation for this and Antimony is provided as markdown files that can be viewed from the application. A summary of the key capabilities is listed in Table~\ref{tbl:Novelty}.

\begin{table}
\centering
\begin{tabular}{llp{8cm}} \toprule 
Feature & Platform & Comment \\ \midrule
Reproducibility & D, W & Export of SBML/Antimony \\
Reproducibility & D & Export of Python scripts, SED-ML and COMBINE archives \\
AI Support & W & The Web version has an embedded AI window that can be used alongside the modeling study. \\
Single window interface & D, W & A single structured window interface makes learning the platform much easier. \\
BioModels interface & D, W & Both versions has fast cached access to the BioModels repository with automatic translation to Antimony syntax. \\
Export Plots as PDF & D, W & Export simulation results in publication-ready PDF files. \\
Data Overlay & D, W & Ability to overlay experimental results onto simulation results. \\
Model Checker & D, W & The Web version has an AI interface that can be used to check for model errors, while the Desktop version has a dedicated model checker built in. \\
Software Innovations: & & Web: Zero install, reduced maintenance requirements \\
& & Desktop: AI-assisted development with future maintenance and new feature additions. \\ \bottomrule
\end{tabular}
\caption{Key Features of the Iridium Platform. D = Desktop version; W = Web Version}
\label{tbl:Novelty}
\end{table}

\section{Use Case: Building a Large EGFR Network}

The unique features of the desktop Iridium application are designed to help in the construction of large-scale models, such as those of the EGFR signaling network. Signaling through the EGFR is mediated through the sequential phosphorylation of downstream substrates. The response of the system can be measured experimentally using quantitation phosphorylation and modeled using mass action-based ODEs. However, parameterizing EGFR network models can be very difficult because of both their size and complexity. For example, our current model that includes only the core kinase signaling pathway between the EGFR and the downstream kinase ERK contains 65 reactions, 58 species and 82 parameters. If the initial parameter estimates are not relatively close to the final values, convergence of parameter estimation runs can be difficult to achieve.

The time needed for convergence to optimal values is highly influenced by how close the initial parameter estimates are to them. We addressed the initial parameter estimation problem by overlaying experimental data on a model segment describing that data and then using the sliders to adjust the appropriate parameters. Regulatory networks, such as the EGFR, will show a characteristic time profile in response to a perturbation. They will also show a characteristic dose-response profile. Thus, experimental data used to parameterize the model included both a time series and a dose series.

To usefully combine data and models, it is essential that the data be normalized to the units used for modeling. In the case of the EGFR network model, these are explicitly laid out in the Antimony code:

\begin{verbatim}
// Unit definitions:
unit substance = 1e-9 mole;
unit time_unit = second;
unit nM = 1e-9 mole / litre;
unit per_nM = litre / 1e-9 mole;
unit per_sec = 1 / second;
unit nM_per_s = 1e-9 mole / (litre * second);
unit per_nMs = litre / (1e-9 mole * second);
\end{verbatim}

The experimental data was normalized to these units by calibrating them against physical standards~\cite{Feng2023.08.03.551714}. The time series data and dose series data contained a common time/dose measurement to facilitate cross-experiment normalization. This improves initial parameter estimates by ensuring quantitative data consistency across different simulation modes.

We use the binding, activation and internalization of the EGFR as an example of how data can be used to generate initial model parameters. The model is a simplified version of the Resat et al.\ model in which the binding of EGF to its receptor both activates the receptor, leading to its self-phosphorylation and accelerated endocytosis~\cite{resat2003integrated}. Directly measured parameters include endocytic rates of the activated receptor, dephosphorylation rates of phospho-EGFR and receptor abundance. Parameters needed to be inferred for the model include the effective forward and reverse rate constant for EGF binding to its receptor and the EGFR self-phosphorylation rate.

To generate data to calibrate the model, EGF at 0.17nM was applied to cells and the phosphorylation level of the EGFR was assessed at 0, 120, 240, 480 and 720 s using mass spectrometry~\cite{Feng2023.08.03.551714}. A dose series in which cells were treated with 0.005nM – 17nM EGF for 240 s was also used. The common point 0.17nM @ 240 s was used as an anchor to normalize the two datasets. The levels of phosphorylated EGFR at each experimental point were quantified and normalized and saved in two text files – one for the time series and one for the dose series. 

\begin{figure}[tbp]
\centering
\includegraphics[scale=1]{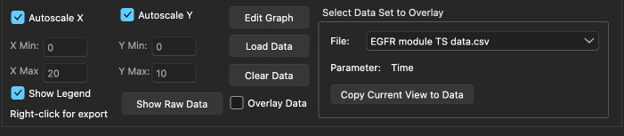}
\caption{Data and graph formatting controls are located below the main graph window. Multiple files can be loaded and switched by the user, but by default the files are associated with the mode in which they were loaded.}
\label{fig:GraphPanel}
\end{figure}

The controls for loading the data overlay are found below the graph itself (Figure~\ref{fig:GraphPanel}). The file should be in standard CSV format with the first row being the labels of the data columns. The title of the first column should be the same as the scanned parameter: “time” in the case of time-series file and “ligand” in the case of the dose series. The color of the default symbols will automatically match that of the simulation result if the labels are the same. Symbols, lines and colors as well as choice of linear or logarithmic graph scaling can be modified using the ``Edit Graph'' button.

\begin{figure}[tbp]
\centering
\includegraphics[scale=0.5]{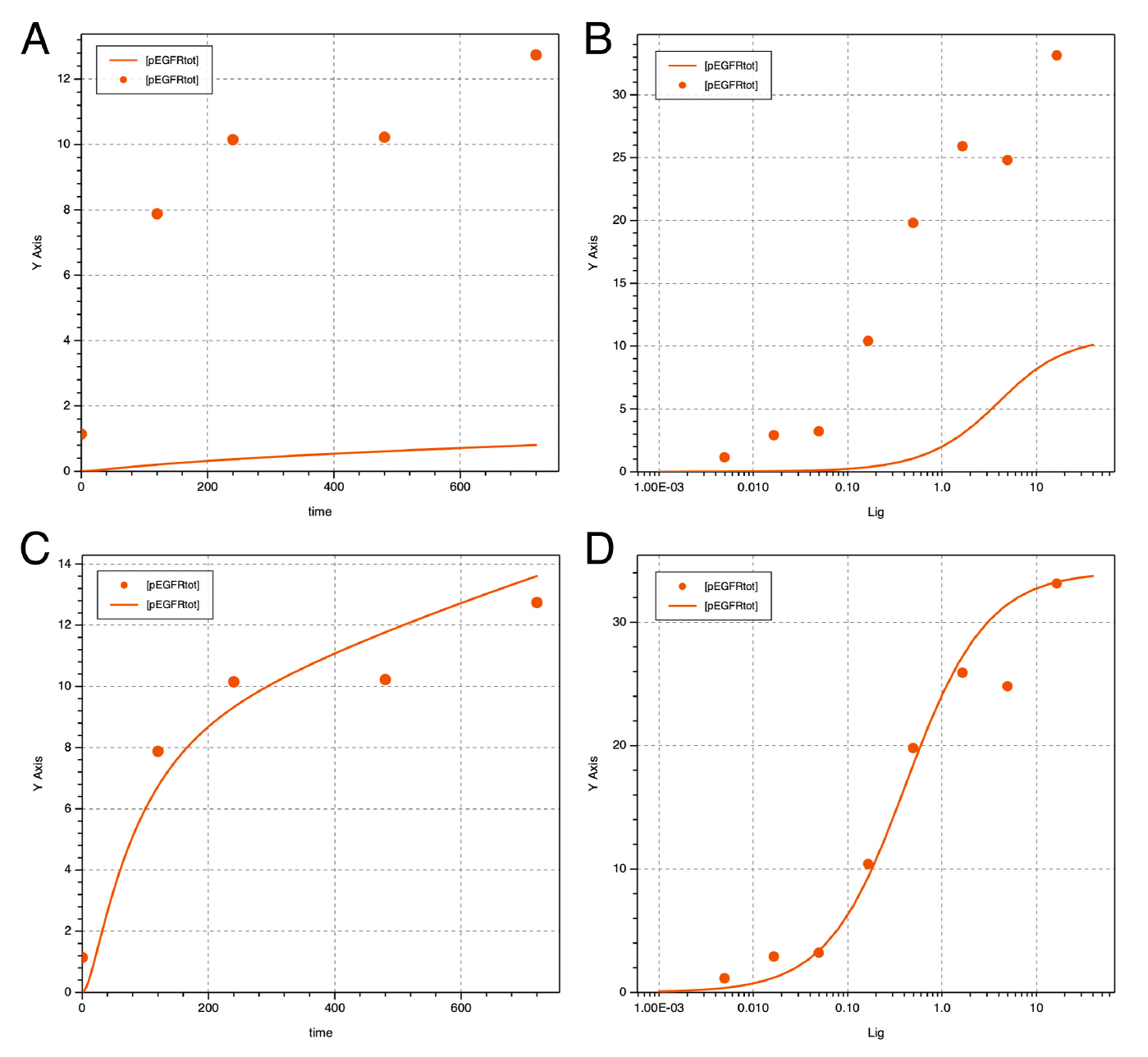}
\caption{Model simulation with data overlay. A: Time Course simulation of the simple EGFR model overlaid with calibrated experimental data. Non-measured initial parameters were taken from the literature. B: Parameter Scan using “ligand” as a scanned boundary species overlaid with calibrated dose-series data using a log scale. Same parameters as in A. C: same as A using parameters interactively set with the sliders. D: same as B using the same adjusted parameters as C.}
\label{fig:FittingPlots}
\end{figure}

The initial EGFR model was constrained by the measured parameters ($ke$, $k_{p1}$, $Vr$, $kt$). Literature values were used for $kf$, $kp1$ and $kr$. As shown in Figure~\ref{fig:FittingPlots}A, correspondence between the simulated and experimental time series results was very poor. Scanning ligand concentration in the Parameter Scan mode showed similarly poor correspondence (Figure~\ref{fig:FittingPlots}B). We then used the sliders to interactively adjust the values of $kf$, $kp1$ and $kr$, initially in the Time Course Simulation mode and then in the Parameter Scan mode. Importantly, the modified parameters were retained when switching modes, allowing a single set of parameters to be optimized using both time-series and dose-series data. 

The model with the adjusted parameters displayed good correspondence with the experimental data (Figure~\ref{fig:FittingPlots}C and D). The slider values were then copied to the model using the button at the bottom left corner of the Time Course Simulation window in preparation for using a parameter estimation program to refine the final model values.

\section{Conclusion}

This article describes the Iridium Platform for simulating biological systems. We describe two Iridium simulation systems. One runs on the desktop (Windows, macOS) and the other is a web-based application that can be run from a simple web hosting site such as GitHub or Cloudflare. The Iridium Platform provides
features that promote ease-of-use by experimental biologists including: sliders for adjusting parameter values while simultaneously displaying simulation results in real time; overlaying experimental data onto simulation results (see use case above); zero-install distribution; AI chat integration; and generation and export of standalone Python implementations that describe the current simulation experiment to support reproducibility.

Both Iridium Simulators are designed around a common workflow involving Control, Input, and Output panels. Model input is provided by the Antimony language, making it very easy to build and edit models. Standard compliance is provided where possible, specifically SBML, SED-ML, and COMBINE archives. 

\begin{table}[htb]
\centering
\begin{tabular}{lp{6.5cm}} \toprule
Enhancement & Comments \\ \midrule
Enzyme kinetics rate laws support. & Remembering specific enzyme kinetic laws is difficult. This enhancement can provide a database of possible enzyme laws that the user can choose from. \\
Stochastic Simulation & Augment deterministic modeling based on differential equations with stochastic simulation. \\
Better SED-ML/SED2 support & Enhance support for computational experiment reproducibility. \\
\bottomrule
\end{tabular}
\caption{Potential enhancements for version 2.0}
\label{tbl:Enhancements}
\end{table}

\section{Availability}

The desktop software is available as binary executables for Windows and macOS. The source code is open source, licensed under the Apache 2.0 license, and available at~\url{https://github.com/sys-bio/IridiumSimulator}. The Web version is available via a static GitHub page and runs on all platforms. The code is open source and licensed under the Apache 2.0 license and is available at \url{https://github.com/sys-bio/WebIridium}

\section*{Acknowledgments}

This work was supported by NIH Biomedical Imaging and Bioengineering award P41 EB023912 through HMS at the Center for Reproducible Biomedical Modeling (\url{https://reproduciblebiomodels.org/}). HSW and HMS were supported by NIH award U01CA227544. SA and HMS were supported by National Institute for General Medical Sciences R01GM155536. FTB was supported by LIBIS and by the Federal Ministry of Research, Technology and Space in the frame of de.NBI \& ELIXIR-DE (W-de.NBI-016). JH and LPS were supported by the reproducibility center, P41 EB023912. The content expressed here is solely the responsibility of the authors and does not necessarily represent the official views of the National Institutes of Health, or any other organization. HMS wrote and designed the desktop software, HSW used the desktop software and drove the feature list, DN wrote the web version. LN wrote the AI interface in the Web version. LPS provided the Antimony bindings for both tools and libroadrunner bindings for the desktop version. JH, HMS, and LPS supervised the development of the Web version. FTB provided the COPASI bindings. SW helped test the GUI and LPS, provided the Antimony functionality. HMS, JH, HSW, FTB, and SA wrote and edited the manuscript. 

%\nolinenumbers

\section*{Appendix}

\section*{Python Script Generated for Figure~\ref{fig:ParameterScan}}

\begin{Verbatim}[fontsize=\small]
# Python script generated by Iridium parameter scan (Figure 3).
# Reproduces the scan using Tellurium.

import tellurium as te
import matplotlib.pyplot as plt
import numpy as np

# -- Model ---------------------------------------------
r = te.loada(r"""
J0: $X0 => S1; VM1*(X0 - S1/Keq1)/(1 + X0 + S1 + S4^h)
J1: S1 => S2; (10*S1 - 2*S2)/(1 + S1 + S2)
J2: S2 => S3; (10*S2 - 2*S3)/(1 + S2 + S3)
J3: S3 => S4; (10*S3 - 2*S4)/(1 + S3 + S4)
J4: S4 => $X1; V4*S4/(KS4 + S4)

// Species initializations:
S1 = 0; S2 = 0; S3 = 0
S4 = 0; X0 = 10; X1 = 0

// Variable initializations:
VM1 = 10; Keq1 = 10
h = 10; V4 = 2.5; KS4 = 0.5

/*
@simulate base: { timeend: 30, points: 500 }

@scan family: {
  source: base,
  parameter: h,
  start: 3,
  end: 8,
  points: 6,
  observables: [S3],
}

@plot: { source: family, y: [S3], title: "Effect of h" }
*/
""")

# -- Selection -----------------------------------------
selection = ['time', '[S3]']

# -- Time course settings ------------------------------
time_start = 0
time_end   = 30
num_points = 500

# -- Scan parameter and range --------------------------
scan_param = 'h'
scan_values = np.linspace(3, 8, 6)

# Colors below match the on-screen plot exactly,
# read from the live series. Edit to restyle.
colors = ['#0000FF', '#1500EA', '#2A00D5', '#3F00C0', '#5500AA', '#6A0095']

# -- Scan ----------------------------------------------
color_idx = 0
for k in scan_values:
    r.reset()
    r.setValue(scan_param, k)
    m = r.simulate(time_start, time_end, num_points, selection)
    for j in range(1, len(selection)):
        c = colors[color_idx] if colors else None
        plt.plot(m[:, 0], m[:, j], color=c,
                 label=f'{selection[j]}, {scan_param}={k:.3g}')
        color_idx += 1
plt.legend()
plt.show()
\end{Verbatim}

\section*{Sample Report from Model Checker}

{\bf\large Model Check Report}

\medskip
2 reaction(s) against 18 rate law(s), structure and behavior.

\section*{What was matched}
\begin{table}[h]
\begin{center}
\begin{tabular}{|l|l|l|}
\hline
Reaction & Rate law & Why\\
\hline
\_J0 & \texttt{michaelis\_menten\_irrev} & a close structural match, distance 0.429 \\
\hline
\_J1 & \texttt{mass\_action\_irrev} & an exact structural match \\
\hline
\end{tabular}
\end{center}
\end{table}

Distance is how far a reaction's expression sits from the law it was matched to: 0.000 is an identical shape, 1.000 nothing in common.

\section*{Findings}

\subsection*{J0 (line 9)}
\textbf{D004 ERROR} (checked against \texttt{michaelis\_menten\_irrev})

this law requires the rate to approach 0.01 as S runs to inf, but at S = 1E010 it is 1E010 and is not closing on it

\begin{itemize}
  \item seen at: \texttt{Km=0.01, Vm=0.01;  S=1E010 gives 1E010, expected \~{}0.01}
\end{itemize}

\textbf{D005 ERROR} (checked against \texttt{michaelis\_menten\_irrev})

this law requires the rate to be 0.005 at S=0.01, but it is 0.01

\begin{itemize}
  \item seen at: \texttt{Km=0.01, S=0.01, Vm=0.01;  rate=0.01, expected 0.005}
\end{itemize}

\textbf{D101 ERROR}

this reaction divides by zero at the values this model starts with (Km = 0), so it cannot be simulated from time zero

\begin{itemize}
  \item found: \texttt{k1*A/Km}
  \item seen at: \texttt{k1=0.35, Km=0, A=10}
\end{itemize}

\textbf{S010 ERROR} (checked against \texttt{michaelis\_menten\_irrev})

the right quantities, but bracketed differently from this law

\begin{itemize}
  \item found: \texttt{k1*A/Km}
  \item expected: \texttt{Vm * S / (Km + S)}
  \item suggestion: \texttt{the law reads Vm * S / (Km + S)}
\end{itemize}

\textbf{D006 WARN} (checked against \texttt{michaelis\_menten\_irrev})

this differs from the law by up to 100.0\% over the sampled range

\begin{itemize}
  \item expected: \texttt{Vm * S / (Km + S)}
  \item seen at: \texttt{Km=0.01, S=1000, Vm=4.64159;  rate 464159 against 4.64154, a 100.0\% difference}
\end{itemize}

\textbf{S011 WARN} (checked against \texttt{michaelis\_menten\_irrev})

``k1'' plays the Vm role, which is conventionally called Vm or Vmax or V\_max or vmax or Vf

\begin{itemize}
  \item found: \texttt{k1}
  \item expected: \texttt{Vm}
\end{itemize}

\subsection*{J1 (line 10)}
\textbf{S018 INFO}

``B'' has no initial value, so it starts at zero

\begin{itemize}
  \item found: \texttt{B}
\end{itemize}

4 errors, 2 warnings.

\section*{Simulation Description Format}

To give users the ability to describe a simulation experiment, we implemented a simple comment-based metadata format. This format is compatible with SED-ML, thus allowing users to share their simulation experiment with other users.

 Simulation metadata is written inside Antimony block comments, so an annotated model remains a valid Antimony file and is ignored by tools that do not recognize the notation. Within a block comment, a command region begins at any line whose first non-whitespace character is @ followed by a command name, and extends to the end of that command's value; text outside a command region is treated as prose and ignored, allowing a descriptive comment and its associated commands to occupy the same block. A command consists of the @ character, a command name, an optional user-supplied label, a colon, and a series of key: value statements enclosed in braces. Values may be numbers, quoted strings, identifiers, booleans, hexadecimal colors, bracketed lists, or nested objects. Whitespace, including newlines, is ignored. key: value pairs are comma-separated. User comments in the form // can still be used to add comments to block. The notation provides no expressions, variables, or control flow. Commands may appear in any number of comment blocks and are processed in file order as though concatenated.

Six commands are defined: {\tt @meta} records title, author and description; {\tt @simulate} and {\tt @steadystate} specify time-course and steady-state tasks; {\tt @scan} repeats a task over a range or explicit list of values of one parameter, reducing each run to a stated measure; and {\tt @plot} and {\tt @output} describe figures and tabular output derived from those tasks. Tasks carry optional labels and can be chained by a source key, so a figure states which run, for example, a simulation will use. A set object on any task assigns values to model quantities such as parameters, species initial values, or compartment sizes, for the duration of that task alone. The parsed representation can be exported as a roadrunner/Tellurium Python script or as SED-ML (Level 1 Versions 3 and 4), optionally packaged as a COMBINE archive that also includes the model. Full documentation for this format is provided as markdown files available within the desktop application itself. The following is a sample file that specifies a time course simulation followed by a plot.

\begin{Verbatim}[fontsize=\small]
$Xo -> S1; Vmax * Km / (Km + S1)
 S1 -> S2; 0.3 * S1
 S2 ->   ; 0.2 * S2

S1 = 1.0;  S2 = 0.5
Vmax = 1.0;  Km = 0.5

/*
  @simulate: { timeend: 50, points: 500 }
  @plot: {
    y: [S1, S2],
    title: "Wild type",
  }
*/
\end{Verbatim}

% This section is obligatory, but may be edited and ammended
\section*{Use of AI}

This project inadvertently served as a use case for using AI to assist in developing sophisticated applications. In this section, we will describe our experience. 

The desktop and web editions of Iridium both use modern AI technology but in different ways. The Web version has a dedicated AI panel where users can ask questions about modeling or even assist in building a model. Currently, it supports Claude and ChatGPT access.

The current desktop version was built entirely using Claude Code~\cite{claudecode2025}. An earlier version of the desktop application exists, and was developed using traditional manual coding techniques. The use of AI to write the second version resulted in a number of key advantages: 1) Development time was much faster, weeks rather than months; 2) Maintenance and bug resolution are now much simpler and handled by AI conversations; 3) Three new GUI Skia components were written from scratch by Claude Code to ensure the functionality was available on macOS. These included the plotting panel, a Markdown viewer, and the syntax highlighting editor. 

\subsection*{Lessons Learned and Observations on the use of AI}

{\bf Specification Document:} A useful lesson learned in using AI for coding was the value of first creating a specification document before attempting to write code. This offered a number of advantages: 

1) It forces a human developer or user to sketch out the basic features required in the application. This allows the AI to propose architectural designs before coding starts.

2) It helps the AI determine the order of construction and it sets up a series of milestones it can follow. This also helps the AI avoid design errors early in the process, resulting in significantly fewer code changes later on. For example, in designing the editor, a redo/undo system is easier to integrate in a later milestone if the AI `knows' that this is a future feature to add and prepare the ground early on; 

3) A specification document allows the AI to plan a more rigorous testing regime for each milestone and will do regression testing on all earlier milestones where applicable. The AI will offload backend logic or algorithms for testing to separate console-based test harnesses. This enables the AI to test specific algorithms without invoking the full GUI application.

4) A specification document can be handed to another developer group, which they can use it to implement the same application in another framework, such as C++ or Rust or even a web-based application. 

For any large project, we therefore recommend creating a specification document first. This practice is well documented by many developers and companies~\cite{piskala2026specdriven} and contrasts with pure `vibe programming', which relies on casual, conversational AI prompting without any planning. In practice, specification-driven coding is more of a hybrid because the specification document is a living document and changes as unanticipated technical constraints emerge or user requirements evolve during implementation. The most important aspect of specification-driven coding is not rigidity but enforced testing and a document that the AI can follow and use to plan with. It also encourages the AI to go into agentic mode, which can result in more robust software.

\noindent{\bf Human Guidance:} While an AI can autonomously write the code without human intervention, the initial architectural design, user requirements, and monitoring of the AI are essential. An AI can sometimes go down `rabbit holes' and needs human intervention to escape and reassess. Often when this happens, it is advisable to start a new session so that it `forgets' the error it just made but can also be told not to follow this type of design in its configuration file.

\noindent{\bf Testing:} The greatest impact on the performance of code written by an AI is by using an agentic tool like Claude Code. This gives the AI the opportunity to obtain as much feedback as possible on how the code performs. This can involve timing code runs and assessing code efficiency, running numerous tests to ensure individual features or subroutines function as expected. For GUI applications, the AI will also, of its own volition, control a GUI application from the keyboard and mouse and take screenshots to assess that the GUI is performing as expected.  For applications that require algorithm development, the AI will create its own separate test harness to ensure the algorithm functions correctly before integrating it into the main application. Each harness gets its own set of tests. An agentic AI will carry out its own testing and bug fixing, which greatly reduces the chance of code `hallucinations'.

\noindent{\bf GUI Development:} Claude Code has proved to be very good at creating graphical user interfaces. For example, the parameter scan panel was designed entirely by the AI, based on prompts from the human developer. However, it is not perfect and required some adjustments. Delphi has a visual designer, so final polishing of the interface is straightforward. 

The final desktop application amounted to just over 98,000 lines of code. Multiple sessions were required for this level of effort because Claude's context window is limited to 1,000,000 tokens. However, Claude Code keeps a record of all its past dealings in the CLAUDE.md file it maintains. A key advantage of using AI has been the reduced cost of maintenance and adding new features, which is now much easier. 

%Total number of lines of code : Core 70015, markdown viewer 11098, plotting component 8047, Antimony editor 8138 Total 97298 lines

\bibliography{refs}

@inproceedings{chance1962analogue,
  title={Analogue and digital computer representations of biochemical processes},
  author={Chance, B and Higgins, JJ and Garfinkel, D},
  booktitle={Federation proceedings},
  volume={21},
  pages={75--86},
  year={1962}
}

@article{fell1992metabolic,
  title={Metabolic control analysis: a survey of its theoretical and experimental development},
  author={Fell, David A},
  journal={Biochemical Journal},
  volume={286},
  number={Pt 2},
  pages={313},
  year={1992}
}

@article{kacser1995control,
  title={The control of flux},
  author={Kacser, Henrik and Burns, James A and Kacser, H and Fell, DA},
  journal={Biochemical Society Transactions},
  volume={23},
  number={2},
  pages={341--366},
  year={1995},
  publisher={Portland Press Ltd.}
}

@misc{claudecode2025,
  author = {{Anthropic}},
  title = {{Claude Code}},
  year = {2025},
  url = {https://claude.ai},
  note = {Command-line AI coding agent}
}

@article{hucka2003systems,
  title={The systems biology markup language (SBML): a medium for representation and exchange of biochemical network models},
  author={Hucka, Michael and Finney, Andrew and Sauro, Herbert M and Bolouri, Hamid and Doyle, John C and Kitano, Hiroaki and Arkin, Adam P and Bornstein, Benjamin J and Bray, Dennis and Cornish-Bowden, Athel and others},
  journal={Bioinformatics},
  volume={19},
  number={4},
  pages={524--531},
  year={2003},
  publisher={Oxford University Press}
}

@article{malik2020biomodels,
  title={BioModels—15 years of sharing computational models in life science},
  author={Malik-Sheriff, Rahuman S and Glont, Mihai and Nguyen, Tung VN and Tiwari, Krishna and Roberts, Matthew G and Xavier, Ashley and Vu, Manh T and Men, Jinghao and Maire, Matthieu and Kananathan, Sarubini and others},
  journal={Nucleic acids research},
  volume={48},
  number={D1},
  pages={D407--D415},
  year={2020},
  publisher = {Oxford University Press}
}

@article{sauro2003next,
  title={Next generation simulation tools: the Systems Biology Workbench and BioSPICE integration},
  author={Sauro, Herbert M and Hucka, Michael and Finney, Andrew and Wellock, Cameron and Bolouri, Hamid and Doyle, John and Kitano, Hiroaki},
  journal={Omics: A journal of integrative biology},
  volume={7},
  number={4},
  pages={355--372},
  year={2003},
  publisher={SAGE Publications Sage CA: Los Angeles, CA}
}

@article{hoops2006copasi,
  title={COPASI—a complex pathway simulator},
  author={Hoops, Stefan and Sahle, Sven and Gauges, Ralph and Lee, Christine and Pahle, J{\"u}rgen and Simus, Natalia and Singhal, Mudita and Xu, Liang and Mendes, Pedro and Kummer, Ursula},
  journal={Bioinformatics},
  volume={22},
  number={24},
  pages={3067--3074},
  year={2006},
  publisher={Oxford University Press}
}

@article{choi2018tellurium,
  title={Tellurium: an extensible python-based modeling environment for systems and synthetic biology},
  author={Choi, Kiri and Medley, J Kyle and K{\"o}nig, Matthias and Stocking, Kaylene and Smith, Lucian and Gu, Stanley and Sauro, Herbert M},
  journal={Biosystems},
  volume={171},
  pages={74--79},
  year={2018},
  publisher={Elsevier}
}

@article{welsh2023libroadrunner,
  title={libRoadRunner 2.0: a high performance SBML simulation and analysis library},
  author={Welsh, Ciaran and Xu, Jin and Smith, Lucian and K{\"o}nig, Matthias and Choi, Kiri and Sauro, Herbert M},
  journal={Bioinformatics},
  volume={39},
  number={1},
  pages={btac770},
  year={2023},
  publisher={Oxford University Press}
}

@article{olivier2005modelling,
  title={Modelling cellular systems with PySCeS},
  author={Olivier, Brett G and Rohwer, Johann M and Hofmeyr, Jan-Hendrik S},
  journal={Bioinformatics},
  volume={21},
  number={4},
  pages={560--561},
  year={2005},
  publisher={Oxford University Press}
}

@article{medley2018tellurium,
  title={Tellurium notebooks—an environment for reproducible dynamical modeling in systems biology},
  author={Medley, J Kyle and Choi, Kiri and K{\"o}nig, Matthias and Smith, Lucian and Gu, Stanley and Hellerstein, Joseph and Sealfon, Stuart C and Sauro, Herbert M},
  journal={PLoS computational biology},
  volume={14},
  number={6},
  pages={e1006220},
  year={2018},
  publisher={Public Library of Science San Francisco, CA USA}
}

@article{sauro1991scamp,
  title={SCAMP: a metabolic simulator and control analysis program},
  author={Sauro, Herbert M and Fell, David A},
  journal={Mathematical and computer modelling},
  volume={15},
  number={12},
  pages={15--28},
  year={1991},
  publisher={Elsevier}
}

@article{chance1943kinetics,
  title={The kinetics of the enzyme-substrate compound of peroxidase},
  author={Chance, Britton},
  journal={Journal of Biological Chemistry},
  volume={151},
  number={2},
  pages={553--577},
  year={1943},
  publisher={Elsevier}
}

@article{chance1952mechanism,
  title={The mechanism of catalase action. II. Electric analog computer studies},
  author={Chance, Britton and Greenstein, David S and Higgins, Joseph and Yang, CC},
  journal={Archives of Biochemistry and Biophysics},
  volume={37},
  number={2},
  pages={322--339},
  year={1952},
  publisher={Elsevier}
}

@article{garny2015opencor,
  title={OpenCOR: a modular and interoperable approach to computational biology},
  author={Garny, Alan and Hunter, Peter J},
  journal={Frontiers in physiology},
  volume={6},
  pages={26},
  year={2015},
  publisher={Frontiers Media SA}
}

@article{piskala2026specdriven,
  author        = {Piskala, Deepak Babu},
  title         = {Spec-Driven Development: From Code to Contract in the Age of AI Coding Assistants},
  journal       = {arXiv preprint arXiv:2602.00180},
  year          = {2026},
  eprint        = {2602.00180},
  archivePrefix = {arXiv},
  primaryClass  = {cs.SE},
  url           = {https://arxiv.org/abs/2602.00180}
}

@inproceedings{Zakai2011Emscripten,
  author    = {Zakai, Alon},
  title     = {Emscripten: an LLVM-to-JavaScript compiler},
  booktitle = {Proceedings of the ACM International Conference Companion on Object Oriented Programming Systems Languages and Applications Companion},
  series    = {OOPSLA '11},
  year      = {2011},
  pages     = {301--312},
  publisher = {Association for Computing Machinery},
  address   = {New York, NY, USA},
  doi       = {10.1145/2048147.2048224},
  url       = {https://doi.org/10.1145/2048147.2048224}
}

@misc{w3cwasm,
  author       = {{World Wide Web Consortium (W3C)}},
  title        = {{WebAssembly Core Specification}},
  url          = {https://www.w3.org/TR/wasm-core-1/},
  year         = {2019},
  note         = {W3C Recommendation. Accessed: 2026-08-11}
}

@article{higgins1964chemical,
  title={A chemical mechanism for oscillation of glycolytic intermediates in yeast cells},
  author={Higgins, Joseph},
  journal={Proceedings of the National Academy of Sciences},
  volume={51},
  number={6},
  pages={989--994},
  year={1964}
}

@article{bergmann2014combine,
  title={COMBINE archive and OMEX format: one file to share all information to reproduce a modeling project},
  author={Bergmann, Frank T and Adams, Richard and Moodie, Stuart and Cooper, Jonathan and Glont, Mihai and Golebiewski, Martin and Hucka, Michael and Laibe, Camille and Miller, Andrew K and Nickerson, David P and others},
  journal={BMC bioinformatics},
  volume={15},
  number={1},
  pages={369},
  year={2014},
  publisher={Springer}
}

@article{waltemath2011reproducible,
  title={Reproducible computational biology experiments with SED-ML-the simulation experiment description markup language},
  author={Waltemath, Dagmar and Adams, Richard and Bergmann, Frank T and Hucka, Michael and Kolpakov, Fedor and Miller, Andrew K and Moraru, Ion I and Nickerson, David and Sahle, Sven and Snoep, Jacky L and others},
  journal={BMC systems biology},
  volume={5},
  pages={1--10},
  year={2011},
  publisher={Springer}
}

@misc{wikipediaLibroadrunnerWikipedia,
	author = {Wikipedia},
	title = {{L}ibroadrunner - {W}ikipedia --- en.wikipedia.org},
	howpublished = {https://en.wikipedia.org/wiki/Libroadrunner},
	year = {2025},
	note = {[Accessed 17-04-2025]},
}

@article{somogyi2015libroadrunner,
  title={libRoadRunner: a high performance SBML simulation and analysis library},
  author={Somogyi, Endre T and Bouteiller, Jean-Marie and Glazier, James A and K{\"o}nig, Matthias and Medley, J Kyle and Swat, Maciej H and Sauro, Herbert M},
  journal={Bioinformatics},
  volume={31},
  number={20},
  pages={3315--3321},
  year={2015},
  publisher={Oxford University Press}
}

@misc{XeroxParc,
    title = {Software Development at Xerox PARC},
    author = {Alan Kay},
    year = {2020},
    howpublished={{https://news.ycombinator.com/item?id=24463842}},
}

@misc{Perl,
    title = {Software Development of Perl},
    author = {Larry Wall},
    year = {2025},
    howpublished = {https://www.amazon.com/b?node=23983471011}
}

@misc{BifurcationGUIApp,
    title = {Interactive Bifurcation tool},
    author = {Herbert M Sauro},
    year = {2026},
    url  = {https://github.com/hsauro/BifurcationTool}
}

@misc{BifurcationKitJulia,
  title = {{BifurcationKit.jl}},
  author = {Veltz, Romain},
  url = {https://hal.archives-ouvertes.fr/hal-02902346},
  institution = {{Inria Sophia-Antipolis}},
  year = {2020},
  month = Jul,
  hal_id = {hal-02902346},
  hal_version = {v1},
  eprint = {hal-02902346},
  eprinttype = {hal},
}

@misc{CAPI,
	author = {Timo Geusch},
	title = {{W}hat does it mean to expose a {C}++ code publicly as a {C} {A}{P}{I} and what are the advantages of doing it? --- softwareengineering.stackexchange.com},
	howpublished = {https://softwareengineering.stackexchange.com/questions/181563/what-does-it-mean-to-expose-a-c-code-publicly-as-a-c-api-and-what-are-the-adva},
	year = {2010},
	note = {[Accessed 16-04-2025]},
}

@article{resat2003integrated,
  title={An integrated model of epidermal growth factor receptor trafficking and signal transduction},
  author={Resat, Haluk and Ewald, Jonathan A and Dixon, David A and Wiley, H Steven},
  journal={Biophysical journal},
  volume={85},
  number={2},
  pages={730--743},
  year={2003},
  publisher={Elsevier}
}

@article {Feng2023.08.03.551714,
	author = {Feng, Song and Sanford, James A. and Weber, Thomas and Hutchinson-Bunch, Chelsea M. and Dakup, Panshak P. and Paurus, Vanessa L. and Attah, Kwame and Sauro, Herbert M. and Qian, Wei-Jun and Wiley, H. Steven},
	title = {A Phosphoproteomics Data Resource for Systems-level Modeling of Kinase Signaling Networks},
	elocation-id = {2023.08.03.551714},
	year = {2023},
	doi = {10.1101/2023.08.03.551714},
	publisher = {Cold Spring Harbor Laboratory},
	URL = {https://www.biorxiv.org/content/early/2023/08/03/2023.08.03.551714},
	eprint = {https://www.biorxiv.org/content/early/2023/08/03/2023.08.03.551714.full.pdf},
	journal = {bioRxiv}
}
\end{document}